\documentclass[letterpaper,12pt]{article}

\usepackage{amssymb,latexsym,amsmath}

\usepackage{fullpage}
\usepackage{nicefrac}
\usepackage{mathrsfs}%different \cal using \mathscr

\usepackage[pdftex]{graphicx}

\makeatletter \@addtoreset{equation}{section}
\makeatother

\begin{document}

\begin{titlepage}
	\thispagestyle{empty}
	\begin{flushright}
		\hfill{DFPD-10/TH/21}
	\end{flushright}
	
	\vspace{35pt}
	
	\begin{center}
	    { \LARGE{\bf Flow equations and attractors for black holes  \\[4mm]  in ${\cal N} = 2$ U(1) gauged supergravity}}
		
		\vspace{50pt}
		
		{Gianguido Dall'Agata \ \& \ Alessandra Gnecchi}
		
		\vspace{25pt}
		
		{
		{\it  Dipartimento di Fisica ``Galileo Galilei''\\
		Universit\`a di Padova, Via Marzolo 8, 35131 Padova, Italy}
		
		\vspace{15pt}
		
		\emph{and}
		
		\vspace{15pt}
		
		{\it   INFN, Sezione di Padova \\
		Via Marzolo 8, 35131 Padova, Italy}}		
		
		\vspace{40pt}
		
		{ABSTRACT}
	\end{center}
	
	\vspace{10pt}
	
We investigate the existence of supersymmetric static dyonic black holes with spherical horizon in the context of ${\cal N} = 2$ U(1) gauged supergravity in four dimensions. We analyze the conditions for their existence and provide the general first-order flow equations driving the scalar fields and the metric warp factors from the asymptotic AdS$_4$ geometry to the horizon. We work in a general duality-symmetric setup, which allows to describe both electric and magnetic gaugings. We also discuss the attractor mechanism and the issue of moduli (de-)stabilization.

\end{titlepage}

\baselineskip 6 mm

%\date{}

%%%%%%%%%%%%%%%%%%%%%%%%%%%%%%%%%%%%%%%%%%%%%%%%%%%%%%%%%%%%%%
%%%%%%%%%%%%%%%%%%%%%%%%%%%%%%%%%%%%%%%%%%%%%%%%%%%%%%%%%%%%%%

\section{Introduction} % (fold)
\label{sec:introduction}

Understanding the microscopic origin of the entropy associated to the black hole horizon area is a primary test for any quantum theory of gravity.
String theory succesfully accomplished this in various instances, though mainly limited to extremal configurations and for asymptotically flat 4- or 5-dimensional black holes.
In this context, supergravity black holes are obtained by configurations of wrapped D-brane states and the microscopic origin of the entropy is understood in terms of state counting in a weakly coupled D-brane setup, which is then extrapolated to the coupling regime where the supergravity approximation can be trusted.

A related aspect of supergravity black holes is the so-called attractor mechanism, which states that the area of the horizon $A_H$ of extremal solutions does not depend on the asymptotic values of the moduli fields, but only on the charges. 
This is a necessary ingredient to ensure a microscopic description of the entropy formula $S = S(A_H)$ by removing the dependence of $S$ on continuous parameters and is valid for generic asymptotically flat configurations.

In this paper we investigate black hole solutions of 4-dimensional $N=2$ gauged supergravity theories, where the matter content is given by vector multiplets and the U(1) gauging is obtained by Fayet--Iliopoulos terms.
The main motivation for considering these toy models is the analysis of the attractor mechanism and of the entropy formula in the case of extremal solutions in theories where there may be a non-trivial cosmological constant and the moduli cannot be freely changed in the solution.
Generically, an Anti--de Sitter (AdS) vacuum stabilizes all the scalar fields and therefore a black hole in AdS may only appear for values of the dilaton such that one cannot extrapolate between strong and weak coupling.

Supersymmetric static black hole solutions in theories with a negative cosmological constant have already been considered in \cite{Sabra:1999ux,Chamseddine:2000bk,Caldarelli:1998hg}, where it was shown that they usually lead to naked singularities, unless higher order derivative corrections are added to the lagrangian.
For this reason, most subsequent approaches to this problem considered extremal non-BPS configurations \cite{Duff:1999gh,Cucu:2003yk,Bellucci:2008cb,Kimura:2010xe}.
One strong limitation of the work in \cite{Sabra:1999ux,Chamseddine:2000bk,Caldarelli:1998hg}, however, was the requirement that the scalar fields remained constant along the solution.
If there is some sort of attractor mechanism at work, the AdS$_4$ vacuum may in fact require a definite value for the scalars that differs from the one required by the construction of a supersymmetric $AdS_2 \times S^2$ horizon geometry.
Hence the appearance of singular geometries.
However, if the scalars are allowed to flow, supersymmetry can be restored and regular geometries can be obtained.
An important step forward in this direction was obtained by the authors of \cite{Cacciatori:2009iz}, who considered a setup like the one of this paper and where supersymmetric black hole configurations were explicitly constructed, though mostly with a hyperbolic horizon.

Although our work uses \cite{Cacciatori:2009iz} as an important basis, we will extend their results in two main directions.
Since the gauging procedure breaks the electric--magnetic duality that a generic 4-dimensional Einstein--Maxwell theory possess, the approach presented in \cite{Cacciatori:2009iz} has the limitation that for the same supergravity model only part of the black hole solutions are accessible, whenever the prepotential defining the scalar $\sigma$-model is fixed.
We will present a completely covariant approach by considering a general U(1) gauged supergravity, where also magnetic gaugings are allowed.
We are also going to describe the black hole solutions by means of first order flow equations driven by a superpotential $W$, which is a function of the scalar fields and the warp factors.
This clearly mimics the flow equations of black holes in ungauged supergravity, where the superpotential is the absolute value of the central charge for supersymmetric configurations \cite{Ferrara:1996dd} or a duality invariant function for non-supersymmetric extremal configurations \cite{Ceresole:2007wx} and gives both the ADM mass at infinity and the horizon area.
However, the different metric ansatz and the presence of a non-trivial cosmological constant usually forbid a direct relation between $W$ and $S$ and/or the mass of the black hole.
As we will show, the general construction of this superpotential proves a very effective procedure in order to obtain explicit solutions.

Before presenting our results, we would like to introduce one last important motivation to the analysis of black hole solutions to gauged supergravity theories: flux compactifications.
It is well known that flux compactifications provide an efficient tool to address the moduli problem in string compactifications.
Fluxes provide a non-trivial source for a potential in the effective theory, as well as deformations leading to gauged supergravity models (see for instance \cite{Polchinski:1995sm,Michelson:1996pn,Dall'Agata:2001zh}).
It is therefore of vital importance for this scenario to understand if there is still an attractor mechanism at work in the case of black hole configurations in gauged supergravities, because their generation may destabilize the vacuum \cite{Hsu:2006vw,Danielsson:2006jg}.
In fact, the presence of a charged black hole may drive the value of the moduli fields to a new value at the horizon, different from the one obtained by the potential generated by flux compactification and eventually catalyze the production of new vacuum bubbles within the original setup \cite{Green:2006nv}.

We should point out that we expect realistic scenarios of flux compactification to require the presence of hypermultiplets.
This means that our analysis should be extended to the case where also this type of scalars is allowed to acquire a non-trivial profile.
In fact, in contrast with the case of ungauged theories, where hyperscalars are moduli of black hole solutions, in gauged supergravity black holes, the hypermultiplet scalars may be charged and hence actively participate to the solution.
A very interesting development in this direction is given by the work of \cite{Hristov:2010eu}, where the authors constructed new solutions in gauged supergravities with non-trivial hypermultiplets, embedding known solutions to the ungauged theories.
A general treatment in terms of a superpotential, like the one presented here, would be desirable for these cases, too.

We should also mention that supersymmetric black holes in gauged supergravities were also analyzed in \cite{Hubscher}, though this paper focussed on non-abelian configurations.

% section introduction (end)

\section{BPS flow equations for dyonic configurations} % (fold)
\label{sec:bps_flow_equations_for_dyonic_configurations}

\subsection{The setup} % (fold)
\label{sub:the_setup}

We are interested in dyonic black hole solutions of ${\cal N} = 2$ U(1) gauged supergravity.
For this reason we are going to consider supergravity models coupled to $n_V$ vector multiplets, a linear combination of which is going to gauge a U(1) factor via suitable Fayet--Iliopoulos (FI) terms.
The bosonic Lagrangian of this class of models is
\begin{equation}
	{\mathscr L} = \frac{R}{2} - g_{i\bar\jmath}\, \partial_\mu z^i \partial^\mu \bar z^{\bar \jmath} + \frac14\,\hbox{Im}{\cal N}_ {\Lambda\Sigma}\, F^\Lambda_{\mu \nu}\, F^{\Sigma\,\mu\nu} +\frac14\,\hbox{Re}{\cal N}_{\Lambda\Sigma}\, F^{\Lambda}_{\mu\nu}\,\frac{\epsilon^{\mu\nu\rho \sigma}}{2\sqrt{-g}} F^\Sigma_{\rho \sigma} - V_g.
	\label{StartingAction} 
\end{equation}
The index $\Lambda=0, 1,\ldots , n_V$ runs over the $n_V$ vectors of the vector multiplets and the graviphoton, $z^i$ denote the complex scalar fields sitting in the vector multiplets and $V_g$ is the scalar potential of the theory generated by the FI terms.
The scalar fields parameterize a special-K\"ahler $\sigma$-model and all the relevant quantities in the Lagrangian and in the supersymmetry transformations can be written in terms of special geometry (We will mostly use notations and conventions as in \cite{Ceresole:1995ca}, but for the spacetime signature).
The $\sigma$-model metric $g_{i \bar \jmath}(z, \bar z)$ can be derived from the second mixed derivatives of the K\"ahler potential, which in turn is a function of the covariantly holomorphic symplectic sections ${\cal V} \equiv e^{K/2}\left(X^\Lambda(z), F_{\Lambda}(z) \right)$, as follows from
\begin{equation}
	1 = i\,\langle {\cal V}, \overline{\cal V}\rangle,
\end{equation}
where the brackets denote the symplectic scalar product $\langle A, B\rangle = A^T \Omega B = A_{\Lambda} B^\Lambda - A^\Lambda B_{\Lambda}$, where   $\Omega$ is the Sp($2n_v+2$) metric.
The vector kinetic matrix ${\cal N}_{\Lambda\Sigma}(z)$ is then a complex and symmetric function of the scalar fields and the scalar potential
\begin{equation}
	V_g = g^{i\bar\jmath}\,  D_i {\cal L} \overline{D}_{\bar \jmath} \overline{\cal L} - 3 |{\cal L}|^2 \qquad \left(\hbox{where}\quad D_i {\cal L} \equiv \partial_i {\cal L} + 1/2\, \partial_i K \, {\cal L}\right)
	\label{Vg}
\end{equation}
can be obtained in terms of the superpotential
\begin{equation}
	{\cal L} =  \langle {\cal G}, {\cal V}\rangle = e^{K/2} \left(X^\Lambda g_{\Lambda}- F_{\Lambda} g^\Lambda\right),
\end{equation}
where ${\cal G} = (g^\Lambda, g_{\Lambda})$ denote the FI terms.
One should not be confused by the fact that we have introduced both electric and magnetic gaugings because in consistent models the electric-magnetic duality group will always allow one to reduce to the case where only electric gaugings are turned on (i.e.~$g^\Lambda = 0$).
However, this also implies a rotation of the symplectic sections and the choice of a somewhat preferred basis.
We therefore prefer to maintain duality covariance and allow for generic FI terms ${\cal G}$.
Although a full $N=2$ duality covariant action has not been built yet, decisive steps have been taken in this direction\footnote{An outline of the procedure that should be followed to obtain the general action by using the embedding tensor formalism can be found in \cite{Louis:2009xd}.}, especially in the case of supergravity coupled to vector multiplets. 
As shown in \cite{Dall'Agata:2003yr,D'Auria:2004yi}, whenever one introduces magnetic gaugings, tensor fields have to be introduced.
In the case of supergravity coupled to vector multiplets, one has therefore to improve couplings to vector-tensor multiplets.
In \cite{Andrianopoli:2007ep} the authors worked out the supersymmetry transformations and scalar potential for supergravity coupled to vector-tensor multiplets and for a generic gauging, although in the case of vanishing FI terms.
The extension to non-trivial FI terms is, however, straightforward and, taking a pragmatic approach, we will use the action (\ref{StartingAction}) as our starting point, as this is going to be the relevant sector for our solutions because we will always consider vanishing tensor fields anyway.

We seek static dyonic black hole configurations. 
Hence we will consider the metric ansatz
\begin{equation}
	ds^2 = -e^{2 U(r)} dt^2 + e^{-2 U(r)} \left(dr^2 + e^{2 \psi(r)} d \Omega^2\right),
	\label{metricansatz}
\end{equation}
where $d \Omega^2$ is going to be the line element of a 2-sphere for most of the applications considered in this paper and appropriate profiles for the vector fields so that
\begin{equation}
	\int_{S^2} F^\Lambda = 4 \pi p^\Lambda, \qquad \int_{S^2} G_\Lambda = 4 \pi q_\Lambda, \qquad \left(\hbox{with} \quad G_{\Lambda} = \frac{\delta {\mathscr L}}{\delta F^\Lambda}\right)
	\label{FG} 
\end{equation}
where $Q \equiv (p^\Lambda,q_\Lambda)$ are the black hole magnetic and electric charges, respectively.
We also assume that the scalar fields have only a radial dependence $z^i = z^i(r)$.
Although we look for static configurations and preserve an SO(3) isometry group along the solutions, the metric ansatz (\ref{metricansatz}) differs from the one of asymptotically flat static configurations because of the additional factor depending on $\psi(r)$.
We inserted this additional factor, because, as we will see, it will be necessary to compensate for the additional curvature contributions to the Einstein equations coming from the (varying) non-trivial cosmological constant.

Once we plug these ansatze in the action (\ref{StartingAction}) we obtain an effective 1-dimensional theory for the scalar fields and the warp factors $U(r)$ and $\psi(r)$
\begin{equation}
	S_{1d} = \int dr\left\{e^{2 \psi}\left[(U^\prime - \psi^\prime)^2 + 2 \psi'{}^2 + g_{i\bar \jmath}z^i{}' \bar z^{\bar \jmath}{}'+ e^{2U-4 \psi} V_{BH} + e^{-2U} V_g+ 2 \psi'' - U''\right] - 1\right\},
	\label{1dS}
\end{equation}
which, after an integration by parts, can be written as
\begin{equation}
	\begin{array}{rcl}
	S_{1d} &=&\displaystyle \int dr\left\{e^{2 \psi}\left[U'{}^2 - \psi'{}^2  + g_{i\bar \jmath}z^i{}' \bar z^{\bar \jmath}{}'+ e^{2U-4 \psi} V_{BH} + e^{-2U} V_g\right] - 1\right\}\\[5mm]
	&&\displaystyle + \int dr \frac{d}{dr}\left[e^{2 \psi}(2\psi'- U')\right].
	\end{array}
\end{equation}
Primes denote derivatives with respect to the radial coordinate and the black hole potential 
\begin{equation}
	V_{BH} = |D {\cal Z}|^2 + |{\cal Z}|^2
\end{equation}
is a function of the central charge
\begin{equation}
	{\cal Z} \equiv \langle Q, {\cal V}\rangle.
\end{equation}
It is also useful to rewrite the black hole potential as
\begin{equation}
	V_{BH} = -\frac12\, Q^T {\cal M} Q,
\end{equation}
where ${\cal M}=(^{AB}_{CD})$ is the symplectic matrix defined by the entries 
\begin{equation}
	\begin{array}{rcl}
		A & = & {\rm Im} {\cal N} + {\rm Re} {\cal N}({\rm Im} {\cal N})^{-1}{\rm Re} {\cal N}, \\[2mm]
		D & = & ({\rm Im} {\cal N})^{-1}, \\[2mm]
		B & = & C^T = -{\rm Re}{\cal N} ({\rm Im} {\cal N})^{-1},\\[2mm]
	\end{array}
	\label{entries} 
\end{equation}

% subsection the_setup (end)

\subsection{BPS rewriting of the action} % (fold)
\label{sub:bps_rewriting_of_the_action}

Since we are interested in analyzing supersymmetric configurations, we have to impose the vanishing of the supersymmetry transformation rules on our background, in addition to solving the equations of motion.
This analysis was performed in this way for generic half-supersymmetric configurations in \cite{Cacciatori:2008ek} and applied to a black hole similar to ours in \cite{Cacciatori:2009iz}, though only for electric gaugings.
The resulting first order differential equations provide solutions to both the supersymmetry conditions and the equations of motion.
We will now extend this work for configurations obtained in the duality-symmetric setup given by (\ref{1dS}).

As a first step in this process, we will show that one can rewrite the action (\ref{1dS}) as a sum of squares of first-order differential equations as long as a specific constraint between the black hole charges and the FI parameters is satisfied.
This rewriting then guarantees the solution of the equations of motion of the effective action.
An important outcome of this rewriting is the existence of an additional constraint on the field configurations that may lead to consistent BPS solutions, which will be identified with the defining equation for a phase factor $\alpha(r)$.
In the next section we will then show how the first-order equations derived here follow from a real superpotential, which is the norm of a complex quantity whose phase is $\alpha$.
A direct analysis of the supersymmetry transformations gives the same results, so we leave the  details of such a derivation for the Appendix.

Following a strategy similar to the one used in the ungauged BPS case in \cite{Denef:2000nb}, we can rewrite the action (\ref{1dS}) as a sum of BPS squares by using a series of special geometry identities.
In particular, we can use the negative-definite matrix ${\cal M}$ as a ``metric'' for a set of symplectic covariant first-order equations.
In order to do so, we will use several special geometry identities.
A basic identity, which will be repeatedly used, is 
\begin{equation}
	\frac12 \left({\cal M} - i \Omega\right) = \Omega \overline {\cal V} \; {\cal V} \Omega + \Omega U_i \, g^{i \bar \jmath}\, \overline{U}_{\bar \jmath} \Omega\,,
\end{equation}
which leads to
\begin{equation}
	{\cal M} {\cal V} = i\, \Omega {\cal V}, \qquad {\cal M} U_i = - i \Omega U_i,
\end{equation}
from which follows that
\begin{equation}
	\overline {\cal V}^{T}{\mathcal M} {\cal V}= i \langle\overline {\cal V}, {\cal V}\rangle =-1\,
\end{equation}
and
\begin{equation}
	U_i^T {\cal M} \overline U_{\bar \jmath} = i \langle U_i , \overline U_{\bar \jmath}\rangle = - g_{i \bar \jmath}.
\end{equation}

The first step is to rewrite the kinetic term for the scalar fields and the scalar potentials $V_g$ and $V_{BH}$ in terms of symplectic sections using
\begin{equation}\label{identities1}
-{\cal V}'^{T} {\cal M} \overline {\cal V}'=g_{i\bar\jmath}z^{i\,\prime}{\bar z}^{\bar\jmath\,\prime}+{\cal A}_r^{2}\,,
\end{equation}
where 
\begin{equation}
	{\cal A}_r \equiv \frac{i}{2}\, \left(\bar z^{\bar \jmath}{}'\, \overline{\partial}_{\bar\jmath}K - z^i{}'\, \partial_i K\right)
\end{equation}
is a composite connection.
Given the properties of the symplectic sections, we can also introduce a phase factor, which we will see related to the spinor projector one imposes in order to solve the supersymmetry equations (see the Appendix), so that 
\begin{equation}
		-{\rm Im}(e^{i\alpha}{\cal V}'^{T}){\cal M}{\rm Im}(e^{i\alpha}{\cal V}')=\frac12g_{i\bar\jmath} z^{i\,\prime}{\bar z}^{\bar\jmath\,\prime}+\frac12{\cal A}_r^{2}\ ,
\end{equation}
and once more obtain new identities:
\begin{eqnarray}\label{identities2}
{\rm Re}(e^{i\alpha}{\cal V})^{T}{\cal M} {\rm Re}(e^{i\alpha}{\cal V})={\rm Im}(e^{i\alpha}{\cal V}^{T}){\cal M}{\rm Im}(e^{i\alpha}{\cal V})=-\frac12\ ,\\[2mm]
{\rm Im}(e^{i\alpha}{\cal V}^{T}){\cal M}{\rm Re}(e^{i\alpha}{\cal V})=0\ ,\\[2mm]
{\rm Im}(e^{i\alpha}{\cal V}')={\rm Im}(e^{i\alpha} z^{i\, \prime}U_{i})-{\cal A}_r\,{\rm Re}(e^{i\alpha}{\cal V})\ ,\\[2mm]
{\rm Im}(e^{i\alpha}{\cal V}^{T}){\cal M}\,Q={\rm Re}(e^{i\alpha}{\cal Z})\ ,\quad {\rm Re}(e^{i\alpha}{\cal V}^{T}){\cal M}\,Q=-{\rm Im}(e^{i\alpha}{\cal Z})\ ,\\[2mm]
{\rm Im}(e^{i\alpha}{\cal V}'){\cal M}\,Q=-{\rm Re}(e^{i\alpha}{\cal Z}')+2\, {\cal A}_r\, {\rm Im}(e^{i\alpha}Z)\ .
\end{eqnarray}
After some long, but straightforward manipulations, the action (\ref{1dS}) can then be rewritten as
\begin{equation}\label{BPS2}
\begin{array}{rcl}
S_{1d}&=&\displaystyle\int dr\left\{-\frac12e^{2(U-\psi)}
{\cal E}^T\mathcal M {\cal E} -e^{2\psi}\left[ (\alpha'+\mathcal A_r) + 2 e^{-U} \, {\rm Re}(e^{-i \alpha} {\cal L}) \right]^{2}\right.\\[5mm]
&&\displaystyle-e^{2\psi}\left[\psi'-2e^{-U}\, {\rm Im}(e^{-i \alpha}\mathcal L) \right]^{2}-\left(1+\langle {\cal G},Q\rangle\right)\\[4mm]
&&\displaystyle\left.-2\frac{d\phantom{r}}{dr}\left[e^{2\psi-U}\,{\rm Im}(e^{-i \alpha}\mathcal L) +\,e^{U}\,{\rm Re}(e^{-i \alpha}\mathcal Z)\right]\right\},
\end{array}
\end{equation}
where we introduced
\begin{equation}
	{\cal E}^T \equiv  2e^{2 \psi}\left(e^{-U}{\mathrm{Im}}(e^{-i\alpha}{\cal V})\right)'\,^{T}-e^{2(\psi-U)}{\cal G}^{T}\Omega \mathcal M^{-1}+4e^{-U}(\alpha'+\mathcal A_r){\mathrm{Re}}(e^{-i\alpha}{\cal V})^{T}
	+Q^{T}.
\end{equation}

A simple inspection of (\ref{BPS2}) shows that we succeeded in rewriting the action (\ref{1dS}) as a sum of squares of first order differential conditions and a boundary term provided the charges fulfill the constraint
\begin{equation}
	\langle {\cal G}, Q\rangle = - 1.
	\label{charge_constraint}
\end{equation}
Once this is satisfied we obtain that BPS configurations have to satisfy three sets of equations
\begin{eqnarray}
	{\cal E} &=& 0, \label{E0}\\
	\psi'&=&2\,e^{-U}\,{\rm Im}(e^{-i \alpha}\mathcal L), \label{psiprime}\\
	\alpha'+\mathcal A_r&=& -2e^{-U}\,\,{\rm Re}(e^{-i \alpha}\mathcal L). \label{alphaprime}
\end{eqnarray}
The first set of conditions contains both the flow equations for the scalar field as well as the equation for the warp factor $U$. 
Equation (\ref{psiprime}) describes the evolution of the other warp factor $\psi$.
Finally, (\ref{alphaprime}) gives the condition on the phase $\alpha$.

Some comments are in order here.
First of all, we can see that the first set of equations reduces to the known BPS equations of the ungauged case as presented in \cite{Denef:2000nb} whenever ${\cal G} = 0$ (and then ${\cal L} = 0$).
In such a case, however, we would get an inconsistency from the constraint (\ref{charge_constraint}).
This implies that the BPS configurations we find by solving such a system are solitonic \cite{Cacciatori:2009iz}.
Actually, the BPS rewriting in the ${\cal G} = 0$ case can be achieved by rewriting the second line of (\ref{BPS2}) as a new squared first order equation and a boundary term
\begin{equation}
	- \left(e^{\psi} \psi-1 \right)^2 -\left(2 e^{\psi}\right)',
\end{equation}
which leads to the identification of $e^{\psi(r)} = r$ and hence to reducing the metric ansatz to the known one of the asymptotically flat configurations.
Then we see that the equations we derived are all symplectic covariant or invariant.
This means that once we obtain some solution in a given frame, for a specific choice of charges $Q$ and FI terms ${\cal G}$, we can map it to a different solution for a different set of charges and FI terms related to the original ones by a duality transformation.
We can also compare our BPS equations with those found in \cite{Cacciatori:2009iz} by identifying $b = e^{-i \alpha - U}$ and setting the magnetic FI terms to zero $g^\Lambda = 0$.
The two sets of conditions match and therefore we can also conclude that our BPS conditions imply also the full 4-dimensional equations of motion.
Finally, we would like to point out that the BPS rewriting of the effective action and the derivation of the first order equations (\ref{E0})--(\ref{alphaprime}) can be trivially extended to the case of flat or hyperbolic horizons and yields the same results, but for the charge constraint (\ref{charge_constraint}), which becomes $\langle {\cal G}, Q\rangle = 0$ or $\langle {\cal G},Q\rangle = 1$ in the flat and hyperbolic case, respectively.

% subsection bps_rewriting_of_the_action (end)

\subsection{Superpotentials and flow equations} % (fold)
\label{sub:superpotentials_and_flow_equations}

Although the BPS square rewriting of the effective 1-dimensional action already led to a set of first-order differential equations for the scalar field dependent symplectic sections ${\cal V}$ and the warp factors, we now provide an explicit expression for the resulting flow equations for the actual scalar fields $z^i$.
This rewriting will lead to the identification of a proper superpotential function driving the BPS flow.

The equation (\ref{E0}) is actually a complex symplectic vector of equations whose information can be extracted by appropriate projections with all possible independent sections. 
We first discuss the projections of the BPS equations ${\cal E} = 0$ on the symplectic sections ${\cal V}$ and their derivatives $U_i$ and then pass to the possible contractions with the charges $Q$ and FI terms ${\cal G}$.
From the contraction 
\begin{equation}
	\langle {\cal E}, {\rm Re}(e^{-i \alpha}{\cal V})\rangle = 0
\end{equation}
we obtain the flow equation for the warp factor $U(r)$:
\begin{equation}
	\label{Uprime}
	U' = -e^{U-2 \psi}\, {\rm Re}(e^{-i \alpha}\mathcal Z) +e^{-U}\,{\rm Im}(e^{-i \alpha}\mathcal L).
\end{equation}
The contraction 
\begin{equation}
	\langle {\cal E}, {\rm Im}(e^{-i \alpha}{\cal V})\rangle = 0
\end{equation}
produces once more an equation for the phase
\begin{equation}
	\alpha' + {\cal A}_r = -e^{U-2 \psi}\, {\rm Im}(e^{-i \alpha}\mathcal Z) -e^{-U}\,{\rm Re}(e^{-i \alpha}\mathcal L).
	\label{alphaprime2}
\end{equation}
Finally, the contraction along the covariant derivatives of the sections 
\begin{equation}
		\langle {\cal E}, U_i\rangle = 0
\end{equation}
leads to the scalar fields flow equations
\begin{equation}
	\label{scalarfloweqs}
	z^i{}' = - e^{i \alpha} g^{i \bar \jmath}\left(e^{U-2 \psi}\overline{D}_{\bar \jmath}\overline{\cal Z}+ i\, e^{-U} \overline D_{\bar \jmath} \overline{\cal L}\right).
\end{equation}
Contractions with $Q$ and/or ${\cal G}$ give identities once (\ref{Uprime}), (\ref{alphaprime2}), (\ref{scalarfloweqs}) and (\ref{alphaprime}) are used.
The first thing we notice is that the flow equation for the phase (\ref{alphaprime2}) differs from the one derived directly from the action, namely (\ref{alphaprime}).
Consistency of the two equations then implies the following constraint:
\begin{equation}
	\label{constraint}
	e^{U-2 \psi}\, {\rm Im}(e^{-i \alpha}\mathcal Z) =e^{-U}\,{\rm Re}(e^{-i \alpha}\mathcal L) .
\end{equation}
The constraint arises as a consequence of the fact that in the BPS rewriting we introduced an additional degree of freedom $\alpha(r)$ that was not present in the reduced action.
We can actually rewrite this constraint as an expression that identifies the phase as
\begin{equation}
  e^{2 i \alpha} = \frac{{\cal Z} - i\, e^{2(\psi-U)}{\cal L}}{\overline{\cal Z} + i\, e^{2(\psi-U)}\overline{\cal L}}.
  \label{phaseconstraint}
\end{equation}
We can see that this phase gets identified with the phase of ${\cal Z}$ in the limit where the gauging goes to zero (or, better, $e^{2i \alpha} = e^{2i \phi_{\cal Z}}$; we will come back on this issue later on).
Another interesting remark is that, by using (\ref{phaseconstraint}), it is straightforward to check that the phase equation (\ref{alphaprime2}) is identically satisfied if the BPS equations associated to the scalar fields and to the warp factor are used.

The other important outcome of this analysis is that we can now realize the BPS condition as flow equations for the effective scalar degrees of freedom $U, \psi, z^i$.
Once we define a superpotential
\begin{equation}
	W \equiv e^{U}\, {\rm Re}(e^{-i \alpha}\mathcal Z)  + e^{-U+2 \psi}\,{\rm Im}(e^{-i \alpha}\mathcal L),
\end{equation}
or, by using the phase constraint (\ref{phaseconstraint}),
\begin{equation}
  W = e^U \, |{\cal Z} - i\, e^{2(\psi-U)}{\cal L}|,
\end{equation}
we can rewrite the flow equations as
\begin{eqnarray}
	U' &=&- g^{UU}\, \partial_U W, \\[2mm]
	\psi' &=&- g^{\psi \psi}\,\partial_{\psi} W,\\[2mm]
	z^i{}' &=& - 2 \,{\widetilde g}^{i \bar \jmath} \,\partial_{\bar \jmath} W,
\end{eqnarray}
where $g_{UU} = -g_{\psi \psi} = e^{2 \psi}$, ${\widetilde g}_{i \bar \jmath} = e^{2 \psi}g_{i \bar \jmath}$ and we used the constraint (\ref{constraint}) in the derivation of the last equation.
It is remarkable that $W$ looks precisely like the norm of a complex quantity whose phase is given by $\alpha$ and that it redces to the supersymmetric superpotential for ${\cal G} = 0$.

Although the structure of the flow equations looks rather neat in these variables, for the subsequent discussion it is useful to rewrite them by introducing a different parameterization for the warp factors.
In detail, we can introduce 
\begin{equation}
	A = \psi - U,
\end{equation}
so that the metric ansatz becomes
\begin{equation}
	ds^2 = -e^{2 U(r)} dt^2 + e^{-2 U(r)} dr^2 + e^{2 A(r)} d \Omega^2.
\end{equation}
By using these variables 
\begin{equation}
  W = e^U \, |{\cal Z} - i\, e^{2A}{\cal L}|
\end{equation}
and the flow equations become
\begin{equation}
	\begin{array}{rcl}
		U' &=&- e^{-2(A+U)}\, \left(W - \partial_A W\right), \\[2mm]
		A' &=& e^{-2(A+U)}\, W,\\[2mm]
		z^i{}' &=& - 2e^{-2(A+U)} \,{g}^{i \bar \jmath} \,\partial_{\bar \jmath} W.
	\end{array}
	\label{flows2}
\end{equation}

% subsection superpotentials_and_flow_equations (end)

% section bps_flow_equations_for_dyonic_configurations (end)

\section{Attractors} % (fold)
\label{sec:attractors}

One of the key properties of extremal black hole solutions is the so-called attractor mechanism discovered in \cite{Ferrara:1995ih,Ferrara:1996dd} for static supersymmetric asymptotically flat black holes and later extended to many other different configurations.
We will now show that such an attractor mechanism is at work also for supersymmetric black holes in U(1) gauged supergravity: we will show that one can write the equations defining the value of the scalar fields at the black hole horizon in terms of a set of algebraic conditions on the charges and the symplectic sections.
We stress, that despite formal similarities, the situation is fundamentally different from the one of asymptotically flat solutions.
In fact, AdS$_4$ solutions already fix the asymptotic value of the moduli, which are then driven to the horizon value by the attractor mechanism.
This means that, although the existence of a black hole horizon specifies the values of the moduli fields in terms of the charges, this attractor cannot be reached from a generic point in moduli space because of the asymptotic constraint in terms of the gauging parameters.

\subsection{Near horizon limit} % (fold)
\label{sub:near_horizon_limit}

When approaching the horizon of a supersymmetric extremal black hole we expect the metric (\ref{metricansatz}) to approach that of an AdS$_2 \times S^2$ spacetime:
\begin{equation}
	ds^2 = -\frac{r^2}{R_A^2}dt^2 + \frac{R_A^2}{r^2}dr^2 + R_S^2 (d \theta^2 + \sin^2 \theta \, d \phi^2),
\end{equation}
where $R_S$ and $R_A$ are the radii of the 2-dimensional sphere and of the 2-dimensional Anti-de Sitter spacetime, respectively.
In the framework of the metric ansatz proposed in (\ref{metricansatz}), this is obtained by imposing
\begin{equation}
	U = \log \frac{r}{R_A}, \qquad \hbox{and} \qquad \psi = \log \frac{r R_S}{R_A},
\end{equation}
or, in terms of the alternative variables for the warp factors,
\begin{equation}
	A = \log R_S.
\end{equation}
This means that 
\begin{equation}
	A' = 0  \qquad  \Leftrightarrow \qquad   W = 0
	\label{horcond1}
\end{equation}
at the horizon.
We also expect the scalar fields to be constant $z^i{}'=0$ at the horizon and therefore we expect \begin{equation}
	\partial_i |{\cal Z} - i\, e^{2A}{\cal L}| = 0 \qquad  \Leftrightarrow \qquad    {D}_{i}{\cal Z}- i\, e^{- 2 A}  D_{i} {\cal L} = 0.
	\label{horcond2}
\end{equation}

The attractor equations can then be obtained by using special geometry identities to expand the moduli independent quantity $Q + i \,e^{2A} \, {\cal G}$ and then use the horizon conditions (\ref{horcond2}).
When we multiply from the left the charge combination just mentioned by $\Omega{\cal M} + i$ we get
\begin{equation}
	\Omega  {\cal M} Q + i Q + i\, e^{2A}\,\Omega {\cal M G} -e^{2A} {\cal G} = 2\left({\cal Z} + i\, e^{2A}\,{\cal L}\right)\,\overline{\cal V}+2 \left(\overline{D}_{\bar\imath} \overline{\cal Z} + i\, e^{2A}\,\overline{D}_{\bar \imath} {\cal L}\right) U^{\bar \imath}.
\end{equation}
This is a general expansion valid at any point of the moduli space.
However, at the attractor point the last term vanishes and we therefore obtain that
\begin{equation}
	Q +e^{2A}\, \Omega {\cal MG}  = -2 {\rm Im}(\overline{\cal Z}{\cal V}) + 2 \, e^{2A}\, {\rm Re}(\overline{\cal L}{\cal V}),
\end{equation}
which is the attractor equation. 
Once again, for ${\cal G} = 0$, we can see that it reduces to the known attractor equation $Q = -2 {\rm Im}(\overline{\cal Z}{\cal V})$.
Since this equation only gives the value of the scalar fields at the attractor point, but we also need to fix the value of $A$ in order to obtain the right geometry, one has to supplement the conditions just derived with the $W = 0$ condition, namely
\begin{equation}
	|{\cal Z}  - i\, e^{2A}{\cal L}| = 0.
\end{equation}
Although this is a real condition, it is easy to see that the request that $e^A$ be a real number gives as an outcome that
\begin{equation}
	e^{2 A} = -i\, \frac{{\cal Z}}{{\cal L}} = R_S^2.
\end{equation}
This equation was also derived in \cite{Cacciatori:2009iz}, as a horizon condition.
Summarizing, the BPS attractors in a U(1) gauged supergravity are
\begin{eqnarray}
	Q +e^{2A}\, \Omega {\cal MG}  &=& -2 {\rm Im}(\overline{\cal Z}{\cal V}) + 2 \, e^{2A}\, {\rm Re}(\overline{\cal L}{\cal V}), \label{attractor1}\\[2mm]
	e^{2 A} &=& -i\, \frac{{\cal Z}}{{\cal L}} = R_S^2. \label{attractor2}
\end{eqnarray}
From the last condition we also learn that the phases of the central charge and of the superpotential of the gauging are related at the horizon, so that
\begin{equation}
	\phi_{\cal Z} = \phi_{\cal L} + \frac{\pi}{2}.
\end{equation}
If we plug this information in the definition of the phase factor $\alpha$ we obtain that $e^{2i \alpha} = e^{2 i \phi_{\cal Z}}$
\begin{equation}
	\alpha = \phi_{\cal Z} + k \,\pi, \quad k \in {\mathbb Z},
\end{equation}
at the horizon.
This is an important consistency requirement, in order to obtain spherical horizons, because we can see from inserting the near horizon limits for the warp factors in the flow equations that at the fixed point
\begin{equation}
	e^{-i \alpha}{\cal Z} = - \frac{R_S^2}{2 R_A} <0
\end{equation}
and this is possible only if the phase $\alpha$ at the horizon is identified with $\phi_{\cal Z} + \pi$.
A different attractor equation was proposed in \cite{Cacciatori:2009iz}, which depends only on the moduli fields. 
This equation can be obtained from ours by plugging (\ref{attractor2}) into (\ref{attractor1}), but it looses the information on the horizon area, which instead is governed by (\ref{attractor2}).

Although the attractor equations (\ref{attractor1})--(\ref{attractor2}) are $2 n_V + 4$ conditions for $2 n_V +1$ variables (the $2 n_V$ scalar fields and the warp factor $A$), we can see that not all of them are independent.
In fact, if we contract (\ref{attractor1}) with ${\cal V}$ we obtain an identity and we can therefore argue that it is equivalent to (\ref{horcond2}), which one recovers by contracting (\ref{attractor1}) with $U_i$.
In order to have a spherical horizon these conditions have to be supplemented by the constraint (\ref{charge_constraint}), which can at times overconstrain the system, as we will show in a while.

More information on the attractor point can also be obtained by further contracting the attractor equation (\ref{attractor1}) by the charges of the gauging or of the black hole and by using (\ref{attractor2}).
In the first case we obtain that
\begin{equation}
	e^{-2A} = 2\left(|D_i {\cal L}|^2 - |{\cal L}|^2\right),
	\label{horg}
\end{equation}
while in the second case we get that
\begin{equation}
	e^{2A} = 2\left(|D_i {\cal Z}|^2 - |{\cal Z}|^2\right).
\end{equation}
These equations are very interesting because they can be related to the second symplectic invariant 
\begin{equation}
	I_2(Q) =  |{\cal Z}|^2 -|D_i {\cal Z}|^2 = - \frac12 Q {\cal M}(F)Q,
\end{equation}
where ${\cal M}(F)$ is a matrix constructed using Re $F_{\Lambda \Sigma}$ and Im $F_{\Lambda \Sigma}$ rather than Re ${\cal N}_{\Lambda \Sigma}$ and Im\;${\cal N}_{\Lambda \Sigma}$.
We can also see that if we start from an AdS$_4$ vacuum $D_i {\cal L} = 0$ and we try to obtain a black hole solution by keeping the scalars constant, we get to an immediate contradictory result, because (\ref{horg}) implies that $e^{-2 A }= - 2 |{\cal L}|^2 <0$.
This excludes the possibility of spherical horizons in an asymptotically AdS geometry while keeping scalars fixed and therefore explains the results of \cite{Sabra:1999ux,Chamseddine:2000bk,Caldarelli:1998hg}.
More in general, the second attractor equation (\ref{attractor2}) can also be written as
\begin{equation}
	e^{2A} = - \frac{{\rm Im}(\overline{\cal Z}{\cal L})}{|{\cal L}|^2},
\end{equation}
which, for $D_i {\cal L}=0$, is equivalent to
\begin{equation}
	e^{2A} = \frac12 \frac{\langle{\cal G}, Q\rangle}{|{\cal L}|^2}\,.
\end{equation}
We then see that this is positive only for hyperbolic horizons, while for spherical horizons $\langle{\cal G}, Q\rangle = -1 <0$.

% subsection near_horizon_limit (end)

% section attractors (end)

\section{Examples of dyonic solutions} % (fold)
\label{sec:examples_of_dyonic_solutions}

We now turn to the analysis of the full flow equations and to the construction of explicit solutions, as an example of how the flow equations work and especially of the fact that now we can obtain in a single duality frame all possible black hole solutions for a given gauged supergravity model.
As explained above, in order to have a regular black hole solution in an asymptotically AdS spacetime, the scalar fields have to flow according to the attractor mechanism discussed in the previous section.
We will now analyze a couple of instances where this is required.
Actually, we will first show that there may be models that do not admit at all such flows, because the AdS$_4$ vacua and the AdS$_2 \times S^2$ can never appear simultaneously for any given set of charges.
We will then investigate the STU model, which is known to admit spherical horizons for special values of the charges \cite{Cacciatori:2009iz}.
We will show that we can find such solutions in the standard frame for the prepotential because of our duality-covariant formulation.

\subsection{One modulus case} % (fold)
\label{sub:one_modulus_case}

One of the simplest special K\"ahler moduli spaces is given by the geometry defined by the prepotential
\begin{equation}
	\label{prepo1}
	F = -i\, X^0 X^1\,.
\end{equation}
This space has only one modulus and the $\sigma$-model metric can be obtained from the K\"ahler potential 
\begin{equation}
	K = - \log 2(z + \bar z),
\end{equation}
which requires that Re$z > 0$.
The gauging potential is determined by 
\begin{equation}
	{\cal L} = e^{K/2}\left(g_0 + i\, g^1 + (g_1 + ig^0)z \right),
\end{equation}
which gives a supersymmetric AdS$_4$ extremum at
\begin{equation}
	z = \frac{g_0 g_1 + g^0 g^1 + i\,(g_0 g^0 - g_1 g^1)}{(g_1)^2 + (g^0)^2}.
\end{equation}
This is in the allowed region of the moduli space if and only if
\begin{equation} 
	g_0 g_1 + g^0 g^1 >0.
\end{equation}

For such a simple model the second derivatives of the prepotential (\ref{prepo1}) are constant and therefore the second symplectic invariant $I_2$ is a constant function of the charges at every point of the moduli space:
\begin{equation}
	I_2({\cal G}) =  |{\cal G}|^2 -|D_i {\cal G}|^2 = - \frac12 {\cal G} {\cal M}(F){\cal G} = g_0 g_1 + g^0 g^1.
\end{equation}
Since at the horizon $e^{-2 A}= - I_2 ({\cal G})$, we immediately see that the requirement to have a regular solution would require
\begin{equation}
	g_0 g_1 + g^0 g^1 <0,
\end{equation}
in direct contradiction with the requirement to have a supersymmetric AdS vacuum.
Hence we conclude that for such a model there are no regular spherical black holes with an AdS asymptotic geometry.
This also implies that the AdS$_4$ vacua of this model will not be destabilized by the presence of supersymmetric black holes.

% subsection one_modulus_case (end)

\subsection{The STU model} % (fold)
\label{sub:the_stu_model}

The STU model is defined by various prepotentials, according to the choice of symplectic frame.
Since our formalism is duality covariant, we can fix a symplectic basis where the prepotential has the classic form
\begin{equation}
	F = \frac{X^1 X^2 X^3}{X^0}\,.
\end{equation}
In this basis the K\"ahler potential is
\begin{equation}
	K = - \log [-i (s - \bar s)(t - \bar t)(u - \bar u)],
\end{equation}
where we introduced normal coordinates $s = X^1/X^0$, $t = X^2/X^0$ and $u = X^3/X^0$.
The symplectic vector ${\cal V}$ for such a prepotential is given by
\begin{equation}
	{\cal V} = e^{K/2}\,(1, s, t, u, -s t u, tu, su, st)^T.
\end{equation}
From \cite{Cacciatori:2009iz} we know that the STU model admits spherical horizon solutions for electric gaugings ${\cal G} = (0, g_{\Lambda})$ and magnetic charges $Q = (p^\Lambda, 0)$, but in the symplectic frame defined by the prepotential
\begin{equation}
	F_{CK} = \sqrt{X^0 X^1 X^2 X^3}.
\end{equation}
The K\"ahler potentials of the two models are obviously the same, but the symplectic sections ${\cal V}$ for the square root prepotential $F_{CK}$ are now
\begin{equation}
	{\cal V}_{CK} = e^{K/2}\, (1, -tu,- su, -st, -s t u, s, t, u)^T.
\end{equation}
The two frames are therefore related by a symplectic transformation
\begin{equation}
	\label{transfS}
	S = \left(\begin{array}{cccccccc}
	1 &  &  &  &  &  &  & \\
	 &  &  &  &  & -1 &  & \\
	  &  &  &  &  &  & -1 & \\
   &  &  &  &  &  &  & -1\\
    &  &  &  & 1 &  &  & \\
    & 1 &  &  &  &  &  & \\
    &  & 1 &  &  &  &  & \\
     &  &  & 1 &  &  &  & 
	\end{array}\right)\,,
\end{equation}
so that ${\cal V}_{CK} = S {\cal V}$.
We should stress that such a transformation is an allowed change of frame, but it is not a duality transformation.
In fact, the duality transformations for the STU model are only a subset of the full symplectic group: SU(1,1)$^3 \subset $ Sp(8,${\mathbb R}$). 
Their form can be computed explicitly (see for instance \cite{Behrndt:1996hu}) and the matrix $S$ does not belong to any of their combinations.
However, the effective 1-dimensional model we started from (\ref{1dS}) is fully constructed out of symplectic invariant quantities.
This means that a solution to the model where, for instance, the gauging potential is obtained from ${\cal L}_{CK} = \langle {\cal G}_{CK}, {\cal V}_{CK}\rangle$, can be mapped to a solution of a different system where ${\cal L} = \langle {\cal G}, {\cal V}\rangle$, with ${\cal V}_{CK} = S {\cal V}$ and ${\cal G} = S^{-1} {\cal G}_{CK}$.
Hence we should be able to reproduce solutions with a spherical horizon for our model, with non-trivial gauging charges ${\cal G} = (0,g^1,g^2,g^3,g_0,0,0,0)^T$ and black hole charges $Q = (p^0,0,0,0,0,q_1,q_2,q_3)$.
In our framework, the superpotential for such a model is given by
\begin{equation}
	W = e^{K/2}|q_1 s + q_2 t + q_3 u + p^0 s t u - i e^{2 A} (g_0 -g^1 t u - g^2 s u - g^3 s t)|.
\end{equation}
By using the flow equations we can immediately check that we can consistently fix the axions Re$\,s$ = Re$\,t$ = Re$\,u = 0$ along the whole solution.
For the remaining flow equations we can then use an Ansatz similar to the one proposed in \cite{Cacciatori:2009iz}, namely (where now $z^i = (s,t,u)$)
\begin{equation}
	{\rm Im}\, z^i = \sqrt{\frac{\frac12 \, |\epsilon_{ijk}| H^j H^k}{H^0 H^i}}\,, \quad  \psi = \log (a r^2 + c)\,, \quad U = -\frac14 \log \frac{H^0 H^1 H^2 H^3}{4},
\end{equation}
and
\begin{equation}
	H^I = \frac{\alpha^I r + \beta^I}{a r^2 + c}.
\end{equation}
After some straightforward manipulation, the resulting equations are
\begin{eqnarray}
	2 p^0 &=& e^{2 \psi}(\partial_r H^0 - g_0 (H^0)^2),\\[2mm]
	-2 q_i &=& e^{2 \psi}(\partial_r H^i - g^i (H^i)^2),\\[2mm]
	2\psi' &=& -g_0 H^0 - \sum_i g^i H^i,
\end{eqnarray}
which can be solved in the same way as in \cite{Cacciatori:2009iz}, although we now see that the charges of their configurations have to be mapped to ours with the appropriate signs, related to the transformation matrix (\ref{transfS}).	
Once the Ansatz for $\psi$ and $H^I$ are used in the above equations one gets
\begin{eqnarray}
	&& 2 p^0 = c \alpha^0 - g_0 (\beta^0)^2, \quad a + g_0 \alpha^0 = 0,\\[2mm]
	&&-2 q_i = c \alpha^i - g^i (\beta^i)^2, \quad a + g^i \alpha^i = 0,\\[2mm]
	&&g_0 \beta^0 + g^i \beta^i = 0, \quad 4 a = - g_0 \alpha^0 - g^i \alpha^i.
\end{eqnarray}

If we look at the simplified setup where all scalar fields can be identified $z^i = -i\, y$, i.e.~$g^i = g$ and $q_i = q$, we can solve the previous equations (together with the constraint to have a spherical horizon) for $a = 1$, $\alpha^0 > 0$, $\alpha^i = \alpha >0$, $g_0 >0$, $g^i = g >0$, $\beta^0 = - 3 (g/g_0 \beta)>0$, $\beta^i = \beta <0$, $p^0 <0$, $q_i = q <0$ and
\begin{equation}
	c = \frac12 - 3 g^2 \beta^2 <0.
\end{equation}
Consistency also gives that
\begin{equation}
	\beta = -\frac{1}{2g} \sqrt{1-4 g q}.
\end{equation}
The final outcome is also consistent at the horizon with the result coming from the attractor equations (\ref{attractor1})--(\ref{attractor2}).
In fact, for the simplified scenario considered here, the imaginary part of (\ref{attractor2}) is identically satisfied and the other equations fix uniquely the scalar fields to
\begin{equation}
	y = \sqrt{\frac{g_0}{2g}} \sqrt{\frac{-1 + 6 gq + \sqrt{1- 16 g q + 48 g^2 q^2}}{1- 3 g q}} >0
\end{equation}
and the warp factor to
\begin{equation}
	e^{2A } = \frac14 \sqrt{\frac{1+ 2 (1-4 g q)\sqrt{1 - 16 gq + 48 g^2 q^2}-3  (1-4 g q)^2}{g_0 g^3}},
\end{equation}
which is precisely the value we obtain by taking the limit for $r \to \sqrt{-c/a}$, i.e.~when we  approach the horizon.

Given our framework, however, we can do more than this.
Since our formalism allows for the introduction of arbitrary electric and magnetic charges both for the gauging as well as for the black hole, once we have fixed a solution, like the one above, we can generate new ones by means of duality transformations.
We actually know that the gauging breaks the duality group SU(1,1)$^3$ to a U(1) related to the isometry of the scalar manifold that is gauged by the graviphoton and the 3 vector fields, which couple to the 4 independent charges of the gauging among the 8 parameters ${\cal G}$.
This means, however, that we can still act with this symmetry on the scalar fields and the gauging and black hole charges.
In particular, we could now generate solutions with non-trivial axions, by using the representation of the three U(1) $\subset $ SU(1,1) duality transformations, which act as follows:
\begin{equation}
 	z^i \to \frac{\cos \theta_i \,z^i + \sin \theta_i}{- \sin \theta_i\,z^i + \cos \theta_i}.
\end{equation}
The action on the charges can be then deduced by the corresponding symplectic transformations derived, for instance, in \cite{Behrndt:1996hu}.

% subsection the_stu_model (end)

% section examples_of_dyonic_solutions (end)

%%%%%%%%%%%%%%%%%%%%%%%%%%%%%%%%%%%%%%%%%%%%%%%%%%%%%%%%%%%%%%

\bigskip
\section*{Acknowledgments}

\noindent We would like to thank D.~Cassani, R.~D'Auria, D.~Klemm, F.~Larsen and A.~Tomasiello for stimulating discussions.
This work is supported in part by the ERC Advanced Grant no. 226455, \textit{``Supersymmetry, Quantum Gravity and Gauge Fields''} (\textit{SUPERFIELDS}), by the Fondazione Cariparo Excellence Grant {\em String-derived supergravities with branes and fluxes and their phenomenological implications} and by the European Programme UNILHC (contract PITN-GA-2009-237920).

\bigskip 

\bigskip 

\appendix

\section{Supersymmetry equations} % (fold)
\label{sub:supersymmetry_equations}

In order to explicitly prove that the configurations discussed so far are supersymmetric, we now analyze in detail the supersymmetry variations of ${\cal N} =2$ U(1) gauged supergravity.
For simplicity we will discuss the case without magnetic gauging parameters, but the extension to the full case is straightforward.
Since we used the mostly plus signature, we will have a sign difference every time there is an upper spacetime index.
The relevant variations are then
\begin{eqnarray}
	\label{trsfSUSY}
\delta\psi_{\mu\,A}&=&D_{\mu}\epsilon_{A}-\varepsilon_{AB}\, T^{-}_{\mu\nu}\,\gamma^{\nu}\,\epsilon^{B}-\frac{i}{2}\,{\cal L}\, \delta_{AB}\,\gamma^{\nu}\,\eta_{\mu\nu}\,\epsilon^{B}\ ,\\[2mm]
\delta\lambda^{iA}&=&-i\,\partial_{\mu}z^{i}\,\gamma^{\mu}\,\epsilon^{A}-G^{-i}_{\mu\nu}\,\gamma^{\mu\nu}\,\varepsilon^{AB}\,\epsilon_{B}+\overline{D}^{i}\overline{\cal L}\,\delta^{AB}\,\epsilon_{B}\ ,
\end{eqnarray}
where the covariant derivative is defined as
\begin{equation}
	D_{\mu}\epsilon_{A}\equiv\partial_{\mu}\epsilon_{A}-\frac14\,\omega_\mu^{ab} \gamma_{ab}\epsilon_{A}+\frac{i}{2}\,{\cal A}_{\mu}\epsilon_{A} + g_{\Lambda}\, A_\mu^\Lambda\, \delta_{AC}\varepsilon^{CB}\epsilon_B,
\end{equation}
and ${\cal A}_\mu$ is the composite connection for the K\"ahler transformations:
\begin{equation}
	{\cal A}_\mu \equiv \frac{i}{2}\, \left(\partial_\mu\bar z^{\bar \jmath}\, \overline{\partial}_{\bar\jmath}K - \partial_\mu z^i\, \partial_i K\right).
\end{equation}
We also have that the vector field strengths $F_{\mu\nu}^\Lambda = 2\partial_{[\mu}A_{\nu]}^\Lambda$ appear via their (anti)self--dual combinations
\begin{equation}
	F_{\mu\nu}^- \equiv \frac12 \left(F_{\mu\nu} - \frac{i}{2}\epsilon_{\mu\nu\rho\sigma} F^{\rho \sigma}\right),
\end{equation}
dressed by the scalar fields
\begin{equation}
\label{trsfPieces}
T^{-}_{\mu\nu}=2i \, {\cal I}_{\Lambda\Sigma}\, L^{\Sigma}\, {F}^{\Lambda-}_{\mu\nu}\, \qquad \qquad
G^{-i}_{\mu\nu}=\overline{D}^i\bar L^{\Gamma}\, {\cal I}_{\Gamma\Lambda}\,F^{\Lambda-}_{\mu\nu}.
\end{equation}

The ansatz for the field strengths is
\begin{eqnarray}\label{vectoransatz}
F^{\Lambda}_{tr}&=&\frac{e^{2U-2\psi}}{2} \, ({\cal I}^{-1})^{\Lambda\Sigma}\, \left(  {\cal R}_{\Sigma \Gamma}\,p^{\Gamma}-q_{\Sigma} \right)\ ,\\[4mm]
F^{\Lambda}_{\theta\phi}&=&-\frac12p^{\Lambda}\sin\theta\ ,
\end{eqnarray}
which, in the combinations (\ref{trsfPieces}), reconstruct the central charge ${\cal Z}$ and its derivatives.

Once the metric ansatz (\ref{metricansatz}), the vector field strengths ansatz (\ref{vectoransatz}) and the requirement that the scalar fields depend only on the radial coordinate is used in the supersymmetry transformations above, we should be able to reproduce the flow equations (\ref{E0})--(\ref{alphaprime}) by requiring the existence of some Killing spinors.

The first variation we analyze is the time component of the gravitino $\delta \psi_{t A} = 0$.
This gives the condition
\begin{equation}
	\frac12 e^{2U}U'\gamma^{01}\epsilon_{A}+ \frac12\, A_t^\Lambda g_{\Lambda} \delta_{AC} \varepsilon^{CB} \epsilon_B +\frac {i}{2}\,e^{3U-2\psi}\,{\cal Z}\,\gamma^{1}\varepsilon_{AB}\epsilon^{B}-\frac{i}{2}\,e^U\,{\cal L}\,\delta_{AB}\gamma^{0}\epsilon^{B}=0,
\end{equation}
where we assumed that $\partial_t \epsilon_A = 0$.
Since this equation contains both chiralities of the 4-dimensional supersymmetry parameters, we need to impose a projector condition that relates them.
We can actually identify the required projectors by rewriting the above equation as
\begin{equation}
	 U'\epsilon_{A}=e^{-2U}\,A_t^\Lambda g_{\Lambda} \, \delta_{AC} \,\gamma^1 \gamma^0 \varepsilon^{CB} \epsilon_B +i\,e^{U-2\psi}\,{\cal Z}\,\gamma^{0}\varepsilon_{AB}\epsilon^{B}-i\,e^{-U}{\cal L}\,\delta_{AB}\gamma^{1}\epsilon^{B}.
\end{equation}
If we introduce two distinct projectors relating the spinor components as
\begin{equation}
	\label{proj1}
	\gamma^0 \epsilon_A = i \,e^{i \alpha} \,\varepsilon_{AB} \epsilon^B
\end{equation}
and
\begin{equation}
	\label{proj2}
	\gamma^1 \epsilon_A = e^{i \alpha} \,\delta_{AB} \epsilon^B,
\end{equation}
we can rewrite the $\delta \psi_{t\, A} = 0$ condition as a single differential equation multiplying the same spinor $\epsilon_A$.
This is proved also using 
\begin{equation}
	\gamma^0 \epsilon^A = -i e^{-i \alpha} \varepsilon^{AB} \epsilon_B \quad \hbox{and} \quad \gamma^1 \epsilon^A = e^{-i \alpha} \delta^{AB} \epsilon_B,
\end{equation}
which follow from (\ref{proj1})--(\ref{proj2}) by consistency.
The resulting time component of the gravitino variation gives
\begin{equation}
	\left(-U'+i e^{-2U}\, A_t^\Lambda g_{\Lambda}-\,e^{U-2\psi}\,e^{-i \alpha}{\cal Z}\,-i\,e^{-U}\, e^{-i \alpha}{\cal L}\right)\epsilon_{A}=0,
\end{equation}
which is satisfied only if the quantity within brackets vanishes.
Identifying the real and imaginary parts of the resulting differential equation, one gets that
\begin{equation}
	\label{susyU}
	U' = - \, e^{U-2 \psi}\, {\rm Re}(e^{-i \alpha} {\cal Z})+e^{-U}{\rm Im}(e^{-i \alpha}{\cal L}) 
\end{equation}
and
\begin{equation}
	\label{susycon1}
	 e^U\, A_t^\Lambda g_{\Lambda}=e^{-U}{\rm Re}(e^{-i \alpha}{\cal L}) +e^{U-2 \psi} {\rm Im}(e^{-i \alpha}{\cal Z}).
\end{equation}

We can now analyze the radial component of the gravitino variation $\delta \psi_{r A} = 0$, which gives
\begin{equation}
\partial_{r}\epsilon_{A}+\frac i2  {\cal A}_r\epsilon_{A} -\frac i{2R^{2}}e^{U-2\psi} {\cal Z}\gamma^{0}\varepsilon_{AB}\epsilon^{B}-\frac i2{\cal  L}\,\delta_{AB}\gamma^{1}e^{-U}\epsilon^{B}=0.
\end{equation}
By using the projectors (\ref{proj1})--(\ref{proj2}) and the supersymmetry conditions (\ref{susyU})--(\ref{susycon1}), this reduces to
\begin{equation}
	\partial_{r}\epsilon_{A}-\frac12 \, \left(U' -  i \widetilde {\cal A}\right)\epsilon_{A} =0,
\end{equation}
where we introduced
\begin{equation}
	\widetilde{\cal A} = {\cal A}_r + \left(e^{U-2 \psi}\,{\rm Im}(e^{-i \alpha} {\cal Z}) + e^{-U}\, {\rm Re}(e^{-i \alpha})\right).
\end{equation}
This equation is readily solved by 
\begin{equation}
	\label{epsradial}
	\epsilon_A = e^{\frac{U}{2} - \frac{i}{2}\int \widetilde{\cal A} \,dr}\chi_A,
\end{equation}
for a spinor $\chi_A$ that is $r$ independent.
Consistency with the projector conditions defined above also imply that
\begin{equation}
	\alpha +\int \widetilde{\cal A} \, dr = 0
\end{equation}
and hence
\begin{equation}
	\alpha'+{\cal A}_r = - e^{U-2 \psi} \, {\rm Im} (e^{-i \alpha}{\cal Z}) - e^{-U}\, {\rm Re}(e^{-i \alpha}{\cal L}),
\end{equation}
reproducing the phase equation (\ref{alphaprime2}).

We are then left with the angular components of the gravitino variations and the dilatino.
From the $\theta$ direction we get that
\begin{equation}
	\partial_{\theta}\epsilon_{A}-\frac12\, e^{\psi}(U'-\psi')\gamma^{12}\epsilon_{A}-\frac12\,e^{U-\psi}\,{\cal Z}\,
	\gamma^{3}\varepsilon_{AB}\epsilon^{B}-\frac i2\,e^{-U+\psi}\,
	{\cal L}\,\delta_{AB}\gamma^{2}\epsilon^{B}=0.
\end{equation}
Once more, using the projectors above as well as the supersymmetry conditions derived so far, we can simplify this equation to
\begin{equation}
	\partial_{\theta}\epsilon_{A} = \frac12\, e^\psi\,\left[\psi'-2 e^U\,{\rm Im}(e^{-i \alpha}{\cal L})+i\left(e^{U-2 \psi}\,{\rm Im}(e^{-i \alpha} {\cal Z})-e^{-U}\,{\rm Re}(e^{-i \alpha}{\cal L})\right)\right]\gamma^{21} \epsilon_A.
\end{equation}
Since the radial dependence is fixed on both sides of the equation by (\ref{epsradial}), we need to require that both the real and imaginary parts of the quantities between square brackets vanish.
This leads to the flow equation for $\psi$ 
\begin{equation}
	\psi'=2  e^U\,{\rm Im}(e^{-i \alpha}{\cal L})
\end{equation}
and to the constraint
\begin{equation}
	e^{U-2 \psi}\,{\rm Im}(e^{-i \alpha} {\cal Z})=e^{-U}\,{\rm Re}(e^{-i \alpha}{\cal L})
\end{equation}
This condition now fixes the ansatz for the time component of the vector fields
\begin{equation}
	A_t^\Lambda g_{\Lambda} = 2\, e^U\, {\rm Re}(e^{-i \alpha}{\cal L}).
\end{equation}
We also get that the Killing spinors $\epsilon_A$ should not depend on $\theta$:
\begin{equation}
	\partial_{\theta} \epsilon_A = 0.
\end{equation}

A similar analysis can be performed for the other angular direction, which gives the same set of flow equations and leaves the following condition on the Killing spinors:
\begin{equation}
	\partial_{\phi}\epsilon_A = \frac12\,\cos \theta\, \gamma^{32} \epsilon_A - \frac i2\, \langle {\cal G}, Q\rangle \, \cos \theta\, \gamma^{01} \epsilon_A.
\end{equation}
This is solved by requiring that
\begin{equation}
	\partial_{\phi}\epsilon_A = 0
\end{equation}
and that 
\begin{equation}
	 \langle {\cal G}, Q\rangle + 1 = 0.
\end{equation}

The only supersymmetry equation remaining is the dilatino variation $\delta \lambda^{iA} = 0$.
By using once more the projector conditions (\ref{proj1})--(\ref{proj2}) and the other supersymmetry constraints obtained above we eventually find the flow equations for the scalar fields:
\begin{equation}
	z^i{}' = - e^{i \alpha} g^{i \bar \jmath}\left[e^{U -2 \psi}\, \overline D_{\bar \jmath} \overline{\cal Z}  + i \, e^{-U}\, \overline D_{\bar \jmath} \overline{\cal L}\right].
\end{equation}

Summarizing, the analysis of the supersymmetry transformations reproduces the flow equations (\ref{E0})--(\ref{alphaprime}) for a Killing spinor of the form
\begin{equation}
	\epsilon_A = e^{\frac{U}{2} + \frac{i}{2}\int \widetilde{\cal A} \,dr}\chi_A,
\end{equation}
where $\chi_A$ is a constant spinor fulfilling
\begin{equation}
	\gamma^0 \chi_A =i\, \varepsilon_{AB} \chi^B, \qquad \gamma^1 \chi_A = \delta_{AB} \chi^B.
\end{equation}
Since we imposed two independent projector conditions, the resulting configurations will be 1/4 BPS (each projector halving the number of preserved supersymmetries).

% subsection supersymmetry_equations (end)

\section{Constant scalar flows} % (fold)
\label{sec:constant_scalar_flows}

As we have explained in the main text, we cannot have regular flows with constant scalars, unless the horizon is not spherical, but for instance hyperbolic \cite{Sabra:1999ux,Chamseddine:2000bk,Caldarelli:1998hg}.
In this case one can have regular solutions by using our flow equations together with the constraint $\langle {\cal G}, Q\rangle = 1$.
If we assume that the scalar fields are fixed at the horizon value, we can impose that
\begin{equation}
	e^{-i \alpha} {\cal Z} = - \frac{R_H^2}{2 R_A}, \quad \hbox{ and } \quad e^{-i \alpha}{\cal L} = \frac{i}{2 R_A}.
\end{equation}
Once inserted in the superpotential we get that
\begin{equation}
	W = \frac{e^U}{2 R_A}\left(e^{2A} - R_H^2\right).
\end{equation}
This implies that the equations for the warp factor reduce to 
\begin{eqnarray}
	U' &=& \frac{e^{-U}}{2 R_A}\left( 1 + R_H^2 e^{-2 A}\right), \\[2mm]
	A' &=& \frac{e^{-U}}{2 R_A}\left( 1 - R_H^2 e^{-2 A}\right).
\end{eqnarray}
A trivial solution is for constant $A$
\begin{equation}
	e^A = R_H, \qquad e^U = \frac{r}{R_A},
\end{equation}
which reproduces the AdS$_2 \times H^2$ horizon solution.
More generally, we can solve these equations first in terms of the variables $A$ and $\psi$, with the equation for $\psi$ being
\begin{equation}
	\label{psip1}
	\psi' = A' + U' = \frac{e^{A-\psi}}{R_A}.
\end{equation}
In fact, introducing now
\begin{equation}
	C = e^{2 A} - R_H^2,
\end{equation}
the differential equations for $A$ and $\psi$ can be used to write
\begin{equation}
	C' = C \psi',
\end{equation}
which is readily solved by
\begin{equation}
	C = k \,e^\psi \quad \Leftrightarrow \quad e^{2A} = R_H^2 + k\, e^\psi,
\end{equation}
where $k = 0$ should give back the AdS$_2 \times H^2$ metric.
Plugging the solution into the equation for $\psi$ (\ref{psip1}), we get that
\begin{equation}
	(e^\psi)' = \frac{\sqrt{R_H^2 + k\, e^\psi}}{R_A},
\end{equation}
which is solved by
\begin{equation}
	e^\psi = k\, \frac{r^2}{4 R_A^2}+ \frac{\sqrt{R_S^2 + k \, \alpha}}{R_A} \,r + \alpha,
\end{equation}
where we chose the integration constant so that the limit $k \to 0$ is well-defined.

If we set $\alpha = 0$, we get that the asymptotic behaviour of the warp factor is
\begin{equation}
	r \to 0: \qquad e^{2 A} \to R_H^2, \quad e^{2U} \to \frac{r^2}{R_A^2},
\end{equation}
which leads to the AdS$_2 \times H^2$ metric
\begin{equation}
	ds^2 = - \frac{r^2}{R_A^2}dt^2 + \frac{R_A^2}{r^2}dr^2 + R_H^2 ds_{H^2}^2,
\end{equation}
and 
\begin{equation}
	r \to \infty: \qquad e^{2 A} \to \frac{k^2}{4 R_A^2}r^2, \quad e^{2U} \to \frac{r^2}{k},
\end{equation}
which leads to a metric that differs from AdS$_4$ by $1/r$ terms in the limit.

% section constant_scalar_flows (end)

%%%%%%%%%%%%%%%%%%%%%%%%%%%%%%%%%%%%%%%%%%%%%%%%%%%%%%%%%%%%%%

\end{document}